# PLASMA CORE ROCKET ENGINES

Gerald E. Marsh

Argonne National Laboratory (Ret)

gemarsh@uchicago.edu

## ABSTRACT

The basic physics for plasma core rocket engines was already completed in 1985. At the time successful containment of a fissioning uranium hexafluoride U(93%)$F_6$ plasma was achieved for ~120 seconds using an argon vortex. Unfortunately, hydrodynamic confinement of a fissioning fuel in a gas core nuclear rocket without uranium loss is probably unachievable. This article proposes a plasma core that is magnetically contained and uses one of many possible stable force-free magnetic field configurations.

**INTRODUCTION**

A recent article by Thomas, et al. [1] proposes a centrifugal nuclear thermal rocket engine that could have specific impulse of ~1800. There is an older concept of plasma core nuclear rocket engine that would have an enormously greater specific impulse.

The basic physics for plasma core rocket engines was already completed in 1985 [2]. At the time successful containment of a fissioning uranium hexafluoride U(93%)F$_6$ plasma was achieved for ~120 seconds using an argon vortex. Unfortunately, hydrodynamic confinement of a fissioning fuel in a gas core nuclear rocket without uranium loss is probably unachievable. Note that U in the context of fissioning plasmas means $^{233}$U or $^{235}$U. At temperatures of interest for a plasma core rocket engine, the UF$_6$ will dissociate and be ionized; at low temperatures, UF$_6$ is a gas. The specific impulse (expressed as effective exhaust velocity, which is used in the older literature) of a plasma core rocket engine could be as high as 50,000 m/s. NASA is currently working on a pulsed plasma rocket engine that could achieve a specific impulse of 5,000 s.

The most attractive geometry for a confined UF$_6$ fissioning plasma would be a sphere. One way of achieving this would be with magnetic confinement rather than hydrodynamic confinement.

One possibility is to use the spheromak configuration [3]. The lowest order spheromak is shown in Fig. 1.

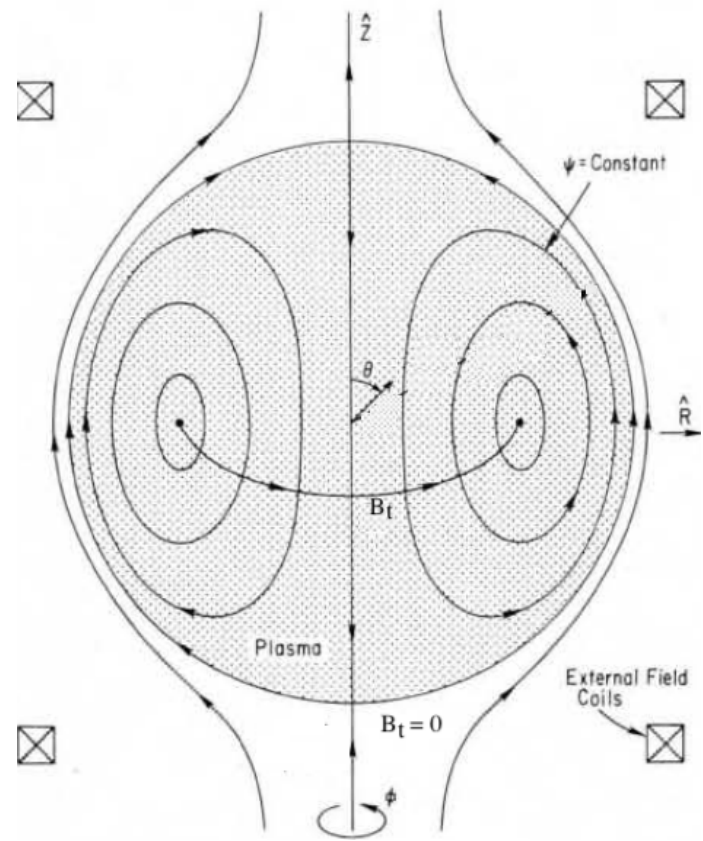

Figure 1. Magnetic configuration of a spheromak. The magnetic field lines form closed magnetic surfaces $\psi(R,Z) = constant$, independent of the cylindrical angle $\phi$. The toroidal field $B_t$ vanishes outside the plasma region. The external field coils provide the longitudinal external magnetic field needed to confine the plasma. [Modified from Fig. 1 of Jardin.]

The spheromak is axially symmetric so that the requirement that $\nabla \cdot B = 0$ means that the poloidal and toroidal parts of the field are given by

$$B_p = (2\pi)^{-1}\nabla\phi \times \nabla\psi(R,Z)$$

and

$$B_t = g(R,Z)\nabla\phi.$$

(Eqs 1).

If the plasma is a non-rotating equilibrium fluid characterized by a pressure *p*, the implication is that *g* and *p* are only a function of $\psi$. For equilibrium one must have

$$J \times B = \nabla p.$$

(Eq 2).

It is generally assumed that the plasma is isotropic. The plasma may not be in complete thermodynamic equilibrium, in which case it is generally assumed that it is in local thermodynamic equilibrium. In that case each elemental volume of the plasma can be characterized by a single temperature. The ideal gas law is then a good equation of state

$$p = nkT = kT \sum_i n_i.$$

(Eq 3).

Here *p* is the total pressure, *n* is the total particle density and $n_i$ is the particle density of the $i_{th}$ species of particle.

If the plasma region in Fig. 1 is considered to be a perfectly conducting fluid the magnetic field lines would be frozen into the plasma so that topological change would be impossible. A small departure from perfect conductivity means that the topology could change. The following question is relevant: If the plasma relaxes to a state that minimizes the magnetic energy with the constraint that helicity be conserved, what is the magnetic field configuration? Woltjer's theorem tells us that the field configuration satisfies [4] the equation

$$\nabla \times B = \alpha B.$$

(Eq 4).

Woltjer considered $\alpha$ to be constant so that this equation describes a linear force-free magnetic field. Taylor reproposed that if the plasma has a small non-zero conductivity the helicity would be approximately invariant so that the minimum energy configuration would still be a linear force-free magnetic field. Equation 2 then tells us that $\nabla p = 0$. This implies that the plasm must have a constant temperature or at least be in local thermodynamic equilibrium. Requiring the temperature variation between the center of the plasma and its surface to be small puts some constraints on the opacity of the plasma. If the opacity is too high there will be a higher temperature in the center of the plasma [5]. One can compensate to some extent by increasing the pressure of the plasma.

Opacity is defined here with respect to the fission fragments of the fissioning $^{235}$U. The fissioning of $^{235}$U is given by $^{235}_{92}U + ^{1}_{0}n \rightarrow ^{141}_{56}Ba + ^{92}_{36}Kr + 3^{1}_{0}n + Energy$. The total energy is about 171 MeV and is carried by the kinetic energy of the fission fragments. The magnitude of the kinetic energy is about the same for each fission fragment [6] and is relatively independent of the energy of the neutron causing the fission.

The problem with the spheromak is that it has serious instabilities so that any perturbation that tilts or shifts the configuration vertically or horizontally will take the spheromak magnetic configuration to a lower energy configuration. Since the minimum energy state is a force-free configuration by Woltjer's theorem [7], one would like a spherical magnetic configuration that is stable. It will be shown below that there probably is one, but where $\alpha$ can be a constant or a function of position.

The classic example of the instability of a non-constant $\alpha$ solution to the force-free field equations is Gold Hoyle nonlinear solution found in 1960 [8]. It represents an infinitely long twisted magnetic field tube that has been proved to be unstable.

This example is not relevant to the solutions discussed below, which being force-free are in a minimal energy configuration so that to change that configuration energy must be gained to remove the field from its force-free configuration to a higher energy state so that it could decay into another force-free configuration different from the original one. The following gives the derivation of the spherical force-free solutions that are of interest [4].

In spherical coordinates ($r$, $\theta$, $\phi$) with the assumption of axial symmetry making the field independent of $\phi$ one can introduce a flux function $\Psi$ such that

$$B_r = \frac{1}{r^2 \sin \theta} \partial_\theta \Psi, \qquad B_\theta = \frac{1}{r \sin \theta} \partial_r \Psi, \qquad B_\phi = \frac{1}{r \sin \theta} f(\Psi),$$

(Eqs. 5).

where $f(\Psi)$ is an arbitrary function of $\Psi$ alone.

One may then show that

$$\partial_r^2 \Psi + \frac{(1-\mu^2)}{r^2} \partial_\mu^2 \Psi = -f(\Psi) f'(\Psi),$$

(Eq. 6).

where $\mu = \cos\theta$ and the prime indicates differentiation by $\Psi$. Also, $\alpha = f'(\Psi)$. Equation 6 is a form of the Grad-Shafranov equation. One can solve this equation by the methos of separation of variables with a separation constant $\lambda$. This results in two equations

$$\Phi''(r) + \left(\alpha^2 - \frac{\lambda}{r^2}\right)\Phi(r) = 0$$

(Eq. 7).

and

$$\Gamma''(\mu) + \frac{\lambda}{(1-\mu^2)} \Gamma(\mu) = 0.$$

(Eq. 8).

The solution to Eq. (8) is given in terms of the generalized Jacobi polynomials $P_m^{(\alpha,\beta)}(\mu)$

$$\Gamma(\mu) = (1-\mu)^{(\alpha+1)/2}(1-\mu)^{(\beta+1)/2} P_m^{(\alpha,\beta)}(\mu)$$

(Eq. 9).

provided Eq. (8) is written as

$$\Gamma''(\mu) + G^{(\alpha,\beta)}(m,\mu) \Gamma(\mu) = 0$$

(Eq. 10).

and $G^{(\alpha,\beta)}(m,\mu)$ is restricted to one of several forms. Apart from an arbitrary multiplicative constant one can obtain for the flux function

$$\Psi = (\alpha r)J_n(\alpha r)(1-\mu)P_n^{(1,-1)}(\mu).$$

(Eq. 11).

The magnetic field components are then given by Eqs. (5) as

$$B_r = \frac{\alpha}{r}J_n(\alpha r)\left\{\frac{n+1}{n(1+\mu)}\left[\mu n P_n^{(1,-1)}(\mu) - (n-1)P_n^{(1,-1)}(\mu)\right],\right\}$$

(Eq. 12).

$$B_\theta = \frac{\alpha(1-\mu)}{r\,sin\theta}P_n^{(1,-1)}(\mu)\{nJ_n(\alpha r) - (\alpha r)J_{n-1}(\alpha r)\},$$

(Eq. 13).

and

$$B_\emptyset = \frac{\alpha}{r\,sin\theta}\left\{B^2 + (\alpha r)^2[J_n(\alpha r)]^2(1-\mu)^2\left[P_n^{(1,-1)}(\mu)\right]^2\right\}^{1/2}.$$

(Eq. 14).

In Eqs. (12-14), instead of $J_n(\alpha r)$ one could have instead chosen $y_n(\alpha r), h_n^1(\alpha r), or\ h_n^2(\alpha r)$, whose properties follow from those of the spherical Bessel functions. These equations provide a very large class of axially symmetric solutions.

Returning to the flux function, for *n* = 1 and particular values of Ψ as a function of *z* = *r* cosθ and *r*, Fig. 2 shows a parametric plot of Ψ.

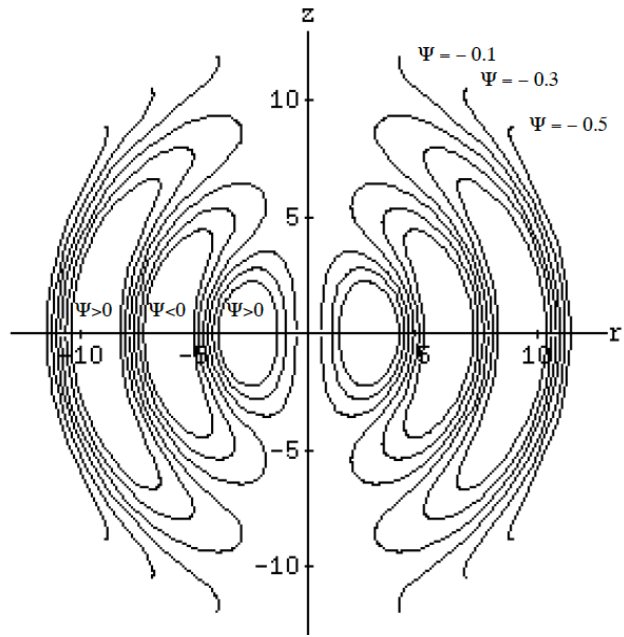

Figure 2. Parametric plot of $\Psi = r\sqrt{\pi/2r}\, J_{3/2}(r)(1-\mu)P_n^{(1,-1)}(\mu)$ as a function of r and $z = r\cos\theta$. (∅ being assumed constant) The contours are for $\Psi = \pm 0.1, \pm 0.3, \pm 0.5$. Note that $\Psi = 0$ on the z-axis.

Combined with Eqs. (5), Fig. (2) allows some insight into the nature of the solutions.

To be used in a plasma core rocket engine one needs to impose a boundary condition on whichever solution is chosen. One possibility is to use external field coils, as in the tokomak shown in Fig 1, to confine the plasma.

Having found a stable force-free magnet field configuration is not sufficient for designing a plasma core rocket engine. One must also find a way to generate the plasma with the field configuration chosen, maintain thermal stability, and find a way to turn the rocket engine off. These are non-trivial engineering problems.

# REFERENCES

[1]. D. Thomas et al., https://doi.org/10.1016/j.actaastro.2025.05.007.

[2]. J.A. Angelo, Jr. and D. Buden, *Space Nuclear Power* [Orbit Book Company, Inc. 1985], Chapter 11, pp. 219-221.

[3]. S.C. Jardin, *Europhysics News*, **17**, Vol 6 (1986).

[4]. G.E. Marsh, *Force-Free Magnetic Fields: Solutions, Topology, and Applications* [World Scientific Publishing Co., 1996].

[5]. K. Thorn and R.T. Schneider, (Eds.), *Research on Uranium Plasmas and their Technological Applications*, Proceedings of a symposium held January 7-8, 1970 in Gainesville, Florida, and sponsored by the National Aeronautics and Space Administration and the University of Florida [NASA, Washington D.C., 1971]; R.W. Patch, *Status of Opacity Calculations for Application to Uranium-Fueled Gas-Core Reactors*, p. 165.

[6] Changqi Liu, et al., "Calculation of the fission products for neutron-induced fission of 235U", *Nuclear Engineering and Technology*, **56,** pp. 1895-1901 (2024).

[7]. L. Woltjer, "A Theorem on Force-Free Magnetic Fields", *Proc. Nat. Acad. of Sci*, **44,** pp. 489-491 **(**1958).

[8]. T. Gold and F. Hoyle, "On the solar flares and the acceleration of cosmic rays" *Monthly Notices of the Royal Astronomical Society,* **120**, pp. 89-104 (1960).